\documentclass[conference]{IEEEtran}
\usepackage[hyphens]{url}
\usepackage{hyperref}
\hypersetup{colorlinks,allcolors=blue}
\usepackage[backend=biber, style=ieee]{biblatex}

\usepackage{enumitem}

\usepackage{amsmath,amssymb,amsfonts}
\usepackage{algorithmic}
\usepackage{graphicx}
\usepackage{multirow}
\usepackage{amsmath}
\usepackage{array}
\usepackage{textcomp}
\usepackage[table,xcdraw]{xcolor}
\usepackage{balance}
\usepackage[flushleft]{threeparttable} 
\usepackage{soul} 
\usepackage{orcidlink}
\usepackage{makecell}

\def\BibTeX{{\rm B\kern-.05em{\sc i\kern-.025em b}\kern-.08em
    T\kern-.1667em\lower.7ex\hbox{E}\kern-.125emX}}

\newcommand{\comments}[1]{}

\begin{document}

\title{Comparing Unidirectional, Bidirectional, and Word2vec Models for Discovering Vulnerabilities in Compiled Lifted Code
}

\author{\IEEEauthorblockN{Gary A. McCully}
  \IEEEauthorblockA{
\textit{Dakota State University}\\
Madison, SD, USA \\
gary.mccully@ieee.org}
\and
\IEEEauthorblockN{John D. Hastings}
\IEEEauthorblockA{
\textit{Dakota State University}\\
Madison, SD, USA \\
john.hastings@dsu.edu}
\and
\IEEEauthorblockN{Shengjie Xu}
\IEEEauthorblockA{
\textit{University of Arizona}\\
Tucson, AZ, USA \\
sjxu@arizona.edu}
\and
\IEEEauthorblockN{Adam Fortier}
\IEEEauthorblockA{
\textit{Georgia Institute of Technology}\\
Atlanta, GA, USA \\
afortier8@gatech.edu}
}

\maketitle

\begin{abstract}
Ransomware and other forms of malware cause significant financial and operational damage to organizations by exploiting long-standing and often difficult-to-detect software vulnerabilities. To detect vulnerabilities such as buffer overflows in compiled code, this research investigates the application of unidirectional transformer-based embeddings, specifically GPT-2. Using a dataset of LLVM functions, we trained a GPT-2 model to generate embeddings, which were subsequently used to build LSTM neural networks to differentiate between vulnerable and non-vulnerable code. Our study reveals that embeddings from the GPT-2 model significantly outperform those from bidirectional models of BERT and RoBERTa, achieving an accuracy of 92.5\% and an F1-score of 89.7\%. LSTM neural networks were developed with both frozen and unfrozen embedding model layers. The model with the highest performance was achieved when the embedding layers were unfrozen. Further, the research finds that, in exploring the impact of different optimizers within this domain, the SGD optimizer demonstrates superior performance over Adam. Overall, these findings reveal important insights into the potential of unidirectional transformer-based approaches in enhancing cybersecurity defenses.
\end{abstract}


\section{Introduction and Background}
Organizations heavily rely upon third-party software to help them achieve their business objectives. For example, countless organizations use Microsoft Windows in their laptop, desktop, and server environments~\cite{RN71}. Microsoft provides these organizations with compiled binaries but does not provide the actual source code.
Due to the loss of high-level structures during compilation from languages such as C/C++, identifying vulnerabilities in compiled binaries is significantly more resource-intensive than in source code. Thus, these organizations must trust the third party (e.g., Microsoft) to follow secure coding practices when developing their applications. Unfortunately, like all software, coding mistakes are made, introducing bugs that threat actors exploit to cause mass destruction to organizations around the world. For example, the WannaCry~\cite{RN75} cryptoworm exploited an SMB vulnerability in Windows systems, encrypted the data on the infected systems, and held the data on these systems for ransom. Recovering from these types of attacks has cost organizations billions of dollars~\cite{RN76}, resulted in economic inflation~\cite{ransomeinfl}, and even cost innocent people their lives~\cite{hospitalfatalities}.

To address the challenge of identifying vulnerabilities within compiled code, some researchers have investigated using machine learning to detect vulnerabilities within compiled code. For example, \cite{RN9} and \cite{schaad2022deeplearningbased} compiled code samples from the NIST Juliet~\cite{RN89} dataset, lifted the compiled code to LLVM using the RetDec~\cite{RN22} tool, and preprocessed the lifted code. This preprocessed code was used to train a word2vec model for creating token-level embeddings, and these embeddings were provided to recurrent neural networks to learn to differentiate between vulnerable and non-vulnerable code.  However, although word2vec is capable of learning high-level code semantics~\cite{mikolov2013efficient}, it lacks the context awareness present in transformer-based Natural Language Processing (NLP) models~\cite{VaswaniSPUJGKP17}. \cite{mccully1} expanded on this research by examining the application of the BERT and RoBERTa models (both bidirectional) to generate embeddings for lifted code. These embeddings were then used to train neural networks to identify stack-based buffer overflows (CWE-121) within the lifted code. The research introduced in this paper builds on the foundational work of \cite{RN9,schaad2022deeplearningbased,mccully1} by comparing and contrasting the impacts of using a unidirectional transformer (GPT-2), bidirectional transformers (BERT, RoBERTa), and non-transformer-based embedding models (Skip-Gram, Continuous Bag of Words) to train an LSTM vulnerability classifier. 

This research is among the first, if not the very first, to present the results of training a GPT model from scratch using LLVM code and using the generated embeddings to locate specific categories of vulnerabilities (e.g., buffer overflows). Furthermore, this research can be compared with the findings in \cite{mccully1} to evaluate which embedding model (BERT, RoBERTa, word2vec, or GPT-2) produces the most effective representation of LLVM code for training a neural network to distinguish between vulnerable and non-vulnerable code. Collectively, the results offer one of the first comparative analyses of neural networks created using embeddings from custom-built unidirectional and bidirectional models, addressing which transformer-based model best represents the LLVM code.

\section{Methodology and Implementation}
\begin{figure}[!htbp]
    \raggedright
    \includegraphics[width=1\linewidth, trim={18 5 17 5}, clip]{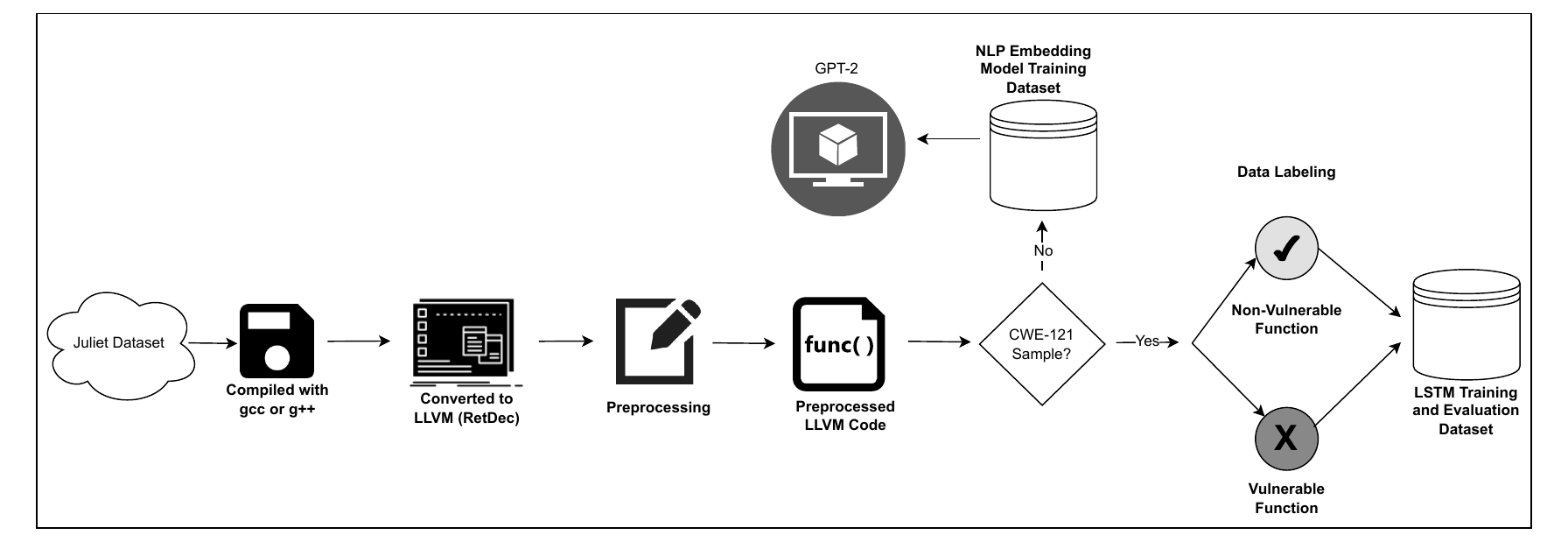}
    \caption{\parbox{\linewidth}{\raggedright Steps to Build GPT-2 Model Using Lifted Code}}
    \label{fig:steps1-4}
\end{figure}

The methodology used in this study involved several steps. As illustrated in Fig. \ref{fig:steps1-4}, the initial steps included:

\begin{enumerate}
\item Juliet dataset samples were compiled into object files.
\item Object files were lifted to LLVM using RetDec.
\item LLVM functions were processed to standardize the code.
\item Samples containing Stack-Based Buffer Overflows (CWE-121) were separated.
\item LLVM samples that did not contain CWE-121 samples were used to train a GPT-2 model.
\end{enumerate}

\begin{figure}[!htbp]
    \includegraphics[width=1\linewidth, trim={18 5 17 5}, clip]{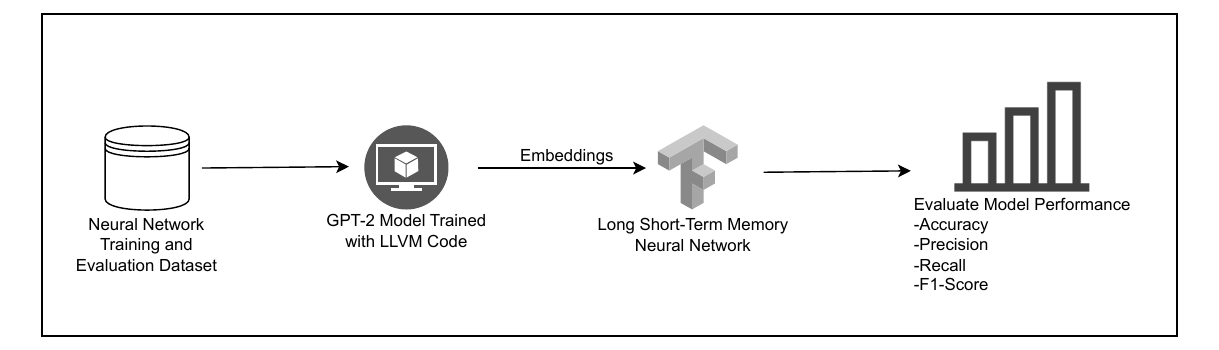}
    \caption{\parbox{\linewidth}{\raggedright Steps for Training a Neural Network to Detect Buffer Overflows}}
    \label{fig:step5}
\end{figure}

As illustrated in Fig. \ref{fig:step5}, the process continued with these steps:

\begin{enumerate}[resume]
\item CWE-121 samples were provided to the GPT-2 model to generate embeddings.
\item Embeddings were used to train long short-term memory (LSTM) neural networks.
\item LSTM performance was evaluated and compared with prior models~\cite{mccully1}.
\end{enumerate}

\noindent The following subsections describe these steps in more detail.

\subsection{Embedding Dataset Creation}\label{embeddingdataset}
The dataset used to train the GPT-2 model was generated by selecting code samples from the NIST SARD Juliet source code dataset without CWE-121 and compiling them on a Kali Linux system\footnote{Originally, the samples were also compiled on a Windows 10 system. However, after inspecting the code lifted to LLVM, it was found that RetDec was stripping calls to the standard C library. Thus, these samples were removed from the dataset.}. The Juliet dataset contains code samples for several vulnerability classes and the corresponding code samples that have implemented a fix to remediate the vulnerability. The code samples were converted into object files on Linux by running the \texttt{make Juliet1.3} command.

The object files were lifted to LLVM using RetDec. The functions of these object files were then isolated, and the preprocessing outlined in \ref{preprocessing} was performed. Although these functions contain a mix of vulnerable and non-vulnerable functions, the dataset was only used to train a GPT-2 model for learning context-aware embeddings. Thus, all functions were placed into a single file without differentiating between vulnerable and non-vulnerable functions. At the end of this process, there were 457,529 functions. However, once duplicates were removed, that number was reduced to 48,157. Table \ref{tab:EmbeddingTrainingDataset} breaks down these numbers.

\subsection{Neural Network Dataset Creation}\label{nndataset}
A subset of the Juliet dataset was used to create the dataset used to train a neural network to differentiate between LLVM functions that contain buffer overflow vulnerabilities and those that do not. As noted previously, the Juliet dataset contains code samples for several vulnerabilities, including CWE-121 (Stack-Based Buffer Overflow). For this research, these code samples were compiled into object files, and the functions were extracted from these files. The Juliet dataset uses a standard naming scheme that indicates functions that contain buffer overflows and those in which the vulnerability was mitigated. The preprocessing steps in \ref{preprocessing} were applied to the data set, but after the preprocessing was completed, it was discovered that some of the functions contained minimal functionality, such as calling a secondary function that contained or did not contain the vulnerability. After preprocessing, most of these functions contained identical instructions and were removed from the dataset when duplicates were eliminated. Finally, functions greater than 2048 tokens in length were removed from the dataset due to resource constraints on neural network training systems. The breakdown of the size of the CWE-121 dataset is given in Table \ref{tab:EmbeddingTrainingDataset}. Given the relatively small dataset, rather than balancing it with an equal number of vulnerable and non-vulnerable functions, metrics such as precision, recall, and F1-score were utilized to evaluate the overall model performance.

\begin{table}[!htbp]
\caption{Embedding and LSTM Training Datasets}
\label{tab:EmbeddingTrainingDataset}
\center
{%
\begin{tabular}{|l|c|c|}
\hline
& \multicolumn{2}{c|}{\textbf{Training (size)}}  \\ \cline{2-3}
\textbf{Dataset Purpose} & LLM & LSTM \\ \hline

\textbf{Object Files	}& 61,023 & 4,949  \\ \hline

\textbf{LLVM Files}	& 61,023 & 4,949 \\ \hline
\textbf{Extracted Functions:}	& & \\

Pre Duplicate Removal	& 457,529 & \makecell{12,069 \\ \Xhline{0.5pt} Clean: 7,011\\ Vuln: 5,058 } \\ \hline

Post Duplicate Removal	& 48,157 & \makecell{3,901\\ \Xhline{0.5pt} Clean: 2,452\\ Vuln: 1,449} \\ \hline

\makecell[l]{Post Removal of Functions\\ More than 2048 Tokens\\} & NA &
\makecell[c]{ 3,802\\ \Xhline{0.5pt} Clean: 2,386\\ Vuln: 1,416 } \\ \hline
\end{tabular}%
}
\end{table}

\subsection{LLVM Preprocessing}\label{preprocessing}
The preprocessing phase removes content unique to each function without affecting how the function works. For example, function names are unique to each function but do not affect how the LLVM code executes. Thus, a function name of \texttt{UniqueFunctionName} can be standardized to \texttt{Func\_one} without losing functionality. The purpose of preprocessing is to remove these aspects of the LLVM functions to standardize how the functions look. The rationale for this choice is the belief that excessive superfluous information within the LLVM functions could hinder the model's ability to learn the most relevant code-level patterns. This study followed the preprocessing steps described in detail in \cite{mccully1}.

\subsection{GPT-2 Model Creation}\label{unidirectionalmodel}
OpenAI in 2018 introduced the first generative pre-training (GPT) model~\cite{radford2018improving}; the idea behind this model is that a base model is built using a large corpus of unlabeled data that can be fine-tuned to perform specific tasks. The second iteration of the GPT model, GPT-2, was introduced shortly after~\cite{radford2019language}. This model was trained on a corpus of 8 million documents and contains 1.5B parameters.

For this study, the GPT2LMHeadModel HuggingFace library was used to create a GPT-2 model that was trained using the same LLVM functions used to train the word2vec, BERT, and RoBERTa models in ~\cite{mccully1}. The system used to train the model had dual TI 4090 GPUs. The tokenizer used to tokenize the LLVM functions was the ByteLevelBPETokenizer tokenizer, and the maximum size of the LLVM functions was set to 2048 tokens. The number of hidden layers was set to 12, the dimensionality of the embeddings was set to 100, and the number of attention heads was set to 10. Finally, the embedding dataset detailed in Section \ref{embeddingdataset} was split into two subsets; one subdataset for training that contained 90\% of the overall dataset, and a second for validation. The model ran for 20 epochs and steadily improved accuracy every 1,000 steps, as demonstrated in Table \ref{tab:GPT2loss}.

\definecolor{lgreen}{rgb}{0.78, 0.99, 0.78} 

\begin{table}[!htbp]
\caption{GPT-2 Five Best Loss Scores by Validation Step}
\label{tab:GPT2loss}
\centering
\begin{tabular}{ccc}
\multicolumn{1}{l}{\textbf{Step}} & \multicolumn{1}{l}{\textbf{Training Set}} & \multicolumn{1}{l}{\textbf{Validation Step}} \\
50,000 & 0.164600 & 0.132824 \\
51,000 & 0.163700 & 0.132133 \\
52,000 & 0.162800 & 0.132173 \\
53,000 & 0.162100 & 0.131591 \\
\cellcolor{lgreen}54,000 & \cellcolor{lgreen}0.163200 & \cellcolor{lgreen}0.131189
\end{tabular}%
\end{table}

\subsection{LSTM Neural Network Configuration}\label{lstmconfig}
Several LSTM neural networks were created. The neural network architecture consisted of a single input layer that contained the GPT-2 embeddings. These embeddings were provided to two hidden LSTM layers with 128 neurons each. The Leaky Rectified Linear Unit (LeakyReLU) was used as the activation function, and a dropout rate of 20\% was used between layers. The output layer was a single dense-layer neuron with a sigmoid activation function used to determine if the function was vulnerable or non-vulnerable. Each neural network was trained for 50 epochs.

The dataset used to train these neural networks is covered in Section \ref{nndataset}. This dataset was split into two subdatasets; the first dataset was 80\% of all data samples and was used to train the model. The second dataset was used for model validation and contained 20\% of the data samples. The same training and validation datasets were used for each model, identical to those used by \cite{mccully1}. When comparing neural networks, this action was taken to remove the random distribution of data samples as a potential variable.

Several LSTM models with different optimizer and optimizer parameters were created for this study. Specifically, neural networks were created using the following hyperparameters
\begin{itemize}
\item Adam optimizer: learning rates of 0.01, 0.001, and 0.0001
\item SGD optimizer: a learning rate and momentum of 0.01
\item SGD optimizer: a learning rate of 0.0001 and momentums of 0.01, 0.001, and 0.0001.
\end{itemize}
Each model was created twice: once with all layers of the embedding model frozen and once with all layers unfrozen.

\section{Results}\label{results}
\subsection{GPT-2 Hyperparameter Comparison}
Table \ref{tab:GPTPerfUnfrozen} details the results of the LSTM training using a GPT-2 model with unfrozen layers. Two-thirds of the neural networks built using the Adam optimizer never achieved accuracy beyond the base accuracy of placing all samples in the same category. However, the models constructed using the SGD optimizer performed much better. The SGD optimizer with an initial learning rate of 0.0001 and a momentum of 0.01 resulted in the best-performing model. This model achieved an accuracy of 92.5\% and an F1-score of 89.7\%.

\begin{table}[!htbp]
\caption{GPT-2 Metrics (Unfrozen LLM Layers)}
\label{tab:GPTPerfUnfrozen}
\resizebox{\columnwidth}{!}{%
\begin{threeparttable}
\begin{tabular}{lcccccc}
\makecell[l]{\textbf{Hyper-}\\ \textbf{parameters}} & \textbf{Epoch} & \textbf{Loss} & \textbf{Accuracy} & \textbf{Precision} & \textbf{Recall} &\makecell[l]{\textbf{F1-} \\ \textbf{Score}} \\
\makecell[l]{SGD\\ -LR: 0.01\\ -Mom: 0.01} & 50 & 0.3049 & 88.6\% & 84.5\% & 84.2\% & 84.4\% \\ \hline
\makecell[l]{SGD\\ -LR: 0.0001\\ -Mom: 0.01} & \cellcolor{lgreen}44 & \cellcolor{lgreen}0.1588 & \cellcolor{lgreen}92.5\% & \cellcolor{lgreen}90.2\% & \cellcolor{lgreen}89.2\% & \cellcolor{lgreen}89.7\% \\ \hline
\makecell[l]{SGD\\ -LR: 0.0001\\ -Mom: 0.001} & 42 & 0.1712 & 92.1\% & 91.6\% & 86.4\% & 88.9\% \\ \hline
\makecell[l]{SGD\\ -LR: 0.0001\\ -Mom: 0.0001} & 46 & 0.1522 & 92.1\% & 91.3\% & 86.7\% & 89.0\% \\ \hline
\makecell[l]{Adam\\ -LR: 0.01} & NA & NA & NA & NA & NA & NA \\ \hline
\makecell[l]{Adam\\ -LR: 0.001} & NA & NA & NA & NA & NA & NA \\ \hline
\makecell[l]{Adam\\ -LR: 0.0001} & 17 & 0.3996 & 87.1\% & 79.1\% & 88.2\% & 83.4\% \\
\end{tabular}%

\begin{tablenotes}
      \small
      \item NA denotes that the model did not improve in accuracy beyond placing all samples in a single category.
    \end{tablenotes}
  \end{threeparttable}
}
\end{table}

In contrast to the performance of neural networks trained with unfrozen embedding models, Table \ref{tab:GPTPerf} presents the results of neural network training using a GPT-2 model with frozen hidden layers. The top-performing neural network trained with frozen embedding model layers was built using the Adam optimizer with a learning rate of 0.001. This neural network achieved an accuracy of 87.9\% and an F1-score of 84.5\%. 

\begin{table}[!htbp]
\caption{GPT-2 Metrics (Frozen LLM Layers)}
\label{tab:GPTPerf}
\resizebox{\columnwidth}{!}{%
\begin{threeparttable}
\begin{tabular}{lllllll}
\begin{tabular}[c]{@{}l@{}}\textbf{Hyper-}\\ \textbf{parameters}\end{tabular} & \textbf{Epoch} & \textbf{Loss} & \textbf{Accuracy} & \textbf{Precision} & \textbf{Recall} & \begin{tabular}[c]{@{}l@{}}\textbf{F1-} \\ \textbf{Score}\end{tabular} \\
\begin{tabular}[c]{@{}l@{}}SGD\\ -LR: 0.01\\ -Mom: 0.01\end{tabular} & 15 & 0.3862 & 82.9\% & 77.9\% & 74.6\% & 76.2\% \\
\hline
\begin{tabular}[c]{@{}l@{}}SGD\\ -LR: 0.0001\\ -Mom: 0.01\end{tabular} & 47 & 0.3210 & 86.9\% & 76.6\% & 92.5\% & 83.8\% \\
\hline
\begin{tabular}[c]{@{}l@{}}SGD\\ -LR: 0.0001\\ -Mom: 0.001\end{tabular} & 42 & 0.3288 & 85.9\% & 76.1\% & 90.0\% & 82.4\% \\
\hline
\begin{tabular}[c]{@{}l@{}}SGD\\ -LR: 0.0001\\ -Mom: 0.0001\end{tabular} & 33 & 0.3615 & 82.8\% & 81.4\% & 68.8\% & 74.6\% \\
\hline
\begin{tabular}[c]{@{}l@{}}Adam\\ -LR: 0.01\end{tabular} & NA & NA & NA & NA & NA & NA \\
\hline
\begin{tabular}[c]{@{}l@{}}Adam\\ -LR: 0.001\end{tabular} & \cellcolor{lgreen}34 & \cellcolor{lgreen}0.2712 & \cellcolor{lgreen}87.9\% & \cellcolor{lgreen}79.7\% & \cellcolor{lgreen}90.0\% & \cellcolor{lgreen}84.5\% \\
\hline
\begin{tabular}[c]{@{}l@{}}Adam\\ -LR: 0.0001\end{tabular} & 40 & 0.2605 & 87.8\% & 77.4\% & 94.3\% & 85.0\% \\
\end{tabular}%

\begin{tablenotes}
      \small
      \item NA denotes that the model did not improve in accuracy beyond placing all samples in a single category.
    \end{tablenotes}
  \end{threeparttable}

}
\end{table}

\subsection{GPT-2 vs. BERT, RoBERTa, CBOW, and Skip-Gram}
The results of \cite{mccully1} were generated using the same process as in the current study. This allows a direct comparison between models trained with GPT-2 embeddings and those using its bidirectional counterparts (e.g., BERT \& RoBERTa). Furthermore, \cite{mccully1} provides results for neural networks trained with simpler word2vec embeddings. Table \ref{tab:llmcompare} integrates those results from \cite{mccully1} and shows that models trained with GPT-2 embeddings outperform their bidirectional counterparts and the simpler word2vec models.

\begin{table}[!htbp]
\caption{Best Performing Neural Networks}
\label{tab:llmcompare}
\resizebox{\columnwidth}{!}{%
\begin{tabular}{lllllll}
\begin{tabular}[c]{@{}l@{}}\textbf{Embedding}\\ \textbf{Model}\end{tabular} & \textbf{Epoch} & \textbf{Loss} & \textbf{Accuracy} & \textbf{Precision} & \textbf{Recall} & \begin{tabular}[c]{@{}l@{}}\textbf{F1-} \\ \textbf{Score}\end{tabular} \\

\begin{tabular}[c]{@{}l@{}}GPT-2\\SGD\\ -LR: 0.0001\\ -Mom: 0.01\end{tabular} & \cellcolor{lgreen}44 & \cellcolor{lgreen}0.1588 & \cellcolor{lgreen}92.5\% & \cellcolor{lgreen}90.2\% & \cellcolor{lgreen}89.2\% & \cellcolor{lgreen}89.7\% \\
\hline

\begin{tabular}[c]{@{}l@{}}word2vec\\ Skip-Gram\\Adam\\-LR: 0.001\end{tabular} & 43 & 0.1848 & 92.0\% & 85.2\% & 94.6\% & 89.6\% \\
\hline

\begin{tabular}[c]{@{}l@{}}BERT\\SGD\\ -LR: 0.0001\\ -Mom: 0.001\end{tabular} & 29 & 0.2625 & 88.8\% & 80.1\% & 92.5\% & 85.9\% \\ \hline

\begin{tabular}[c]{@{}l@{}}RoBERTa\\SGD\\ -LR: 0.0001\\ -Mom: 0.001\end{tabular} & 45 & 0.2708 & 88.8\% & 87.6\% & 81.0\% & 84.2\% \\
\hline

\begin{tabular}[c]{@{}l@{}}word2vec\\ CBOW\\Adam\\-LR: 0.001\end{tabular} & 34 & 0.2528 & 87.5\% & 78.0\% & 91.8\% & 84.3\% \\ \hline
\end{tabular}%
}
\end{table}

\section{Discussion}\label{discussion}
\subsection{Unidirectional vs. Bidirectional Embeddings}
The neural network achieving the highest performance with the GPT-2 embeddings outperformed the best LSTM model utilizing the BERT and RoBERTa embeddings. In this regard, embeddings generated in a single direction, similar to how a human would read code, could better capture code-level semantics than models that attempt to learn semantics from both directions. \cite{devlin2019bert} demonstrates that one of the strengths of BERT is classification tasks. However, the methodology used by this study does not use BERT directly for a classification task; instead, it uses these models to capture code-level semantics. The downstream LSTM is ultimately responsible for classifying the vulnerable code. In contrast, \cite{radford2018improving} shows that GPT models excel at text generation. In this regard, it is possible that the generative characteristics of the GPT model can better represent the semantics of the code than those of BERT in the context of the current study. However, future research would need to be conducted to validate that this is indeed the case.

\subsection{Unidirectional vs. Skip-Gram Embeddings}
In terms of accuracy, the top-performing neural network using GPT-2 embeddings slightly outperformed the best model with Skip-Gram embeddings. However, the Skip-Gram-based model achieved a significantly higher recall score, while the F1-scores for both models were nearly identical. This suggests that the best models using GPT-2 and Skip-Gram embeddings performed similarly. GPT-2 embeddings consistently achieved accuracies around 92\% with the SGD optimizer, a learning rate of 0.0001, and various momentum values. However, models using Skip-Gram embeddings reached this level of accuracy only once, when trained with the Adam optimizer and a learning rate of 0.001~\cite{mccully1}.

\section{Future Work}\label{tab:futurework}
\subsection{Dataset Sizes}
Smaller datasets were used to train and evaluate the GPT-2 (approximately 48K LLVM functions) and LSTM (approximately 4K LLVM functions) models. Although large language models typically require extensive datasets to train properly, the GPT-2 model in this study was trained on a relatively small dataset. Despite the limited data, the study demonstrated strong performance. Increasing the number of training samples would likely improve the quality of the embeddings and better represent the LLVM code. 

\subsection{Data Snooping} 
\citeauthor*{arp2022and} \cite{arp2022and} states that data snooping is one of the biggest pitfalls in research on the application of machine learning to problems in the cybersecurity field. One form of data snooping, test snooping, occurs when training and validation data are not strictly segmented. In the context of the current study, test snooping was not a factor since the samples containing CWE-121 were not used to train the GPT-2 model. However, due to the prominence of the issue within existing research~\cite{arp2022and}, the effects of test snooping on model accuracy were studied in \cite{datasnooping} with no impact found.

\subsection{Embedding Layer Tuning Enhancements}
The research took an all-or-nothing approach to freezing and unfreezing the embedding model layers. \cite{howard2018} highlights the benefits of techniques such as discriminative fine-tuning and gradual unfreezing. Greater model performance may be possible using these methods.

\section{Related Work}
Current academic literature has extensively studied the use of both unidirectional and bidirectional transformer-based large language models (LLMs) for detecting vulnerabilities in source code. These studies include the use of RoBERTa~\cite{10190505, 10507456, 10.1145/3564625.3567985}, BERT~\cite{shestov2024finetuning, liu2024source, 9985089, 10507456, 9484500, 9817336, 10.1145/3564625.3567985}, DistilBERT~\cite{10517582, 10507456, mathews2024, 10.1145/3564625.3567985}, and GPT models~\cite{10224924, zhang2024promptenhanced, du2024generalizationenhanced, wang2024m2cvd, liu2024vuldetectbench, khare2024understanding, zhang2024empirical, 10.1145/3639476.3639762, jones2024codesentry, 10381286, 10301302} to detect vulnerabilities in source code. Furthermore, there have been multiple studies that have focused on the use of bidirectional~\cite{9631481, 10.5555/3615924.3615947, koo2021semanticaware, yu2022firmvulseeker, 10.1145/3564625.3567975} and unidirectional~\cite{10.1145/3460120.3484587} embedding models for code similarity detection. 

Although several papers have employed LLMs to identify vulnerabilities in source code, considerably fewer have focused on using them to detect specific categories of vulnerabilities (e.g., buffer overflows) in compiled code. In \cite{RN67}, researchers train a RoBERTa model from scratch using assembly code instructions. The embeddings from this model are used to train a Message Passing Neural Network (MPNN) to differentiate between vulnerable and non-vulnerable code. \cite{gallagherllvm} also built a RoBERTa model to learn code-level semantics to train a neural network to detect vulnerabilities. However, rather than training the RoBERTa model with assembly code, the researchers trained it with LLVM code. \cite{han2022binary} trained a BERT model using P-Code; these P-Code embeddings were used to train a model that could detect vulnerabilities within the lifted code.

Much of the process around dataset selection, the lifting of the assembly code to LLVM, and the preprocessing of the lifted code within this study is based on the work of \cite{RN9} and \cite{schaad2022deeplearningbased}. Both of these studies used the SARD source code dataset, compiled the samples in this dataset, and lifted the assembly code to LLVM using the RetDec tool. Furthermore, the lifted code was used to train a word2vec model to generate embeddings of the LLVM code, and these embeddings were used to train a neural network to identify vulnerabilities. The work of~\cite{mccully1} expanded on this work by using the LLVM code to train custom BERT and RoBERTa models to generate the LLVM embeddings.

\section{Conclusion}\label{conclusion}
This research aims to provide valuable insight that enables product engineers to create real-world solutions to detect vulnerabilities in compiled binaries. Given the devastating effects of malware on countless organizations worldwide, this research is expected to help the industry make strides in addressing this critical issue.

The authors believe this research represents the first instance of training a unidirectional NLP encoder with LLVM functions to discover specific categories of vulnerabilities in compiled code. Furthermore, when reviewed in the context of~\cite{mccully1}, it is the first study to compare recurrent neural networks created using bidirectional and unidirectional NLP encoders to discover vulnerabilities in compiled code. The results demonstrate that with a smaller training dataset (approximately 48K), embeddings generated from a GPT-2 model are better suited for identifying vulnerabilities in lifted code than those generated from BERT or RoBERTa. Furthermore, this study provides insights into the effects of optimizers on the performance of neural networks built using GPT-2 embeddings and the substantial impact of freezing and unfreezing embedding model layers. Neural networks created with the SGD optimizer outperformed networks built using the Adam optimizer. Furthermore, neural networks trained using GPT-2 embeddings outperformed the simpler word2vec CBOW embeddings and achieved comparable performance to models trained using Skip-Gram embeddings. While GPT-2 embeddings demonstrated higher accuracy and precision, Skip-Gram embeddings achieved a higher recall.
\balance
\printbibliography
\end{document}